\documentclass[fleqn,10pt]{wlscirep}

\usepackage{gensymb}
\usepackage{bm}
\usepackage{times}
\usepackage{graphicx}
\usepackage{array}
\usepackage{float}
\usepackage{colortbl}
\usepackage{caption}
\usepackage{multirow}
\usepackage{tabularx}

\title{Superconductivity of novel tin hydrides (Sn$_n$H$_m$) under  pressure}

\author[1]{M. Mahdi Davari Esfahani}
\author[1,2]{Zhenhai Wang}
\author[1,3,4,*]{Artem R. Oganov}
\author[1]{Huafeng Dong}
\author[1]{Qiang Zhu}
\author[1]{Shengnan Wang}
\author[1]{Maksim S. Rakitin}
\author[1]{Xiang-Feng Zhou}

\affil[1]{Department of Geosciences, Center for Materials by Design, and Institute for Advanced Computational Science,
State University of New York, Stony Brook, NY 11794-2100, USA}
\affil[2]{Peter Grunberg Research Center, Nanjing University of Posts and Telecommunications, Nanjing, Jiangsu 210003, China}
\affil[3]{Skolkovo Institute of Science and Technology, Skolkovo Innovation Center, Bldg.3, Moscow 143026, Russia}
\affil[4]{Department of Problems of Physics and Energetics, Moscow Institute of Physics and Technology, 9 Institutskiy Lane,
Dolgoprudny City, Moscow Region 141700, Russia}
\affil[5]{School of Physics and Key Laboratory of Weak-Light Nonlinear Photonics, Nankai University, Tianjin 300071, China}
\affil[*]{To whom correspondence should be addressed. Email: artem.oganov@stonybrook.edu}

\begin{abstract}
With the motivation of discovering high-temperature superconductors, evolutionary algorithm is employed to search for all stable compounds in the Sn-H  system. 
 In addition to the traditional SnH$_4$, new hydrides SnH$_8$, SnH$_{12}$ and SnH$_{14}$ are found to be thermodynamically stable at high pressure. 
Dynamical stability and superconductivity of tin-hydrides are systematically investigated. 
I$\bar{4}$m2-SnH$_8$, C2/m-SnH$_{12}$ and C2/m-SnH$_{14}$ exhibit higher superconducting transition temperatures of 81, 93 and 97 K compared to the traditional compound SnH$_4$ with T$_c$ of 52 K at 200 GPa. 
An interesting  bent H$_3^-$ in I$\bar{4}$m2-SnH$_8$ and novel liner H$_4^-$ in C2/m-SnH$_{12}$ are observed. 
All the new tin-hydrides remain metallic  over their predicted range of stability. 
The intermediate-frequency wagging and bending vibrations have more contribution to electron-phonon coupling parameter than high-frequency stretching vibrations of H$_2$ and H$_3$.
\end{abstract}

\begin{document}

\flushbottom
\maketitle


%
%
\thispagestyle{empty}

\section*{Introduction}

Molecular hydrogen's phase transition to a metallic state has been  subject of many experimental and theoretical studies \cite{Nellis2006,Loubeyre2002}.
Although reaching the metallic state in pure solid hydrogen proved elusive, it is in the main focus of many groups and recently, the progress of bringing pure hydrogen to nearly 400 GPa has been reported \cite{Zha2013,Howie2012,Review2014}.
Following the pioneering work of Ashcroft \cite{Ashcroft1968}, nearly room-temperature superconductivity was predicted in metallic molecular hydrogen \cite{Cudazzo2010,Cudazzo2010a}.

An alternative approach in reaching  enormous pressure to metalize hydrogen is 
to use chemical alloying as a means to exert additional pressure on hydrogen atoms \cite{Ashcroft2004}. 
 Hydrogen-rich compounds such as SiH$_4$ can be metalized at a much lower pressure \cite{Eremets2008}. For metallic hydrogen, high Debye temperature and strong electron-phonon coupling are anticipated. The same is expected for hydrogen-rich compounds and it has been  suggested that hydrogen-rich compounds are good candidates for high-temperature superconductivity \cite{Ashcroft2004}. Theoretical studies confirm this idea with predicting high-temperature superconductivity in high-pressure hydrides such as H-Se \cite{Zhang}, Ca-H \cite{Wang2012}, Sn-H \cite{Gao2010}, Pt-H \cite{PhysRevB.84.054543} and B-H \cite{Hu2013}. 
A series of hydrogen-rich compounds have been predicted to have remarkably high T$_c$ values (e.g. 235 K for CaH$_6$ at 150 GPa \cite{Wang2012}, 191 K for H$_3$S at 200 GPa \cite{Duan2014}, 64 K for GeH$_4$ at 220 GPa \cite{Gao2008}) 
 while the highest T$_c$ that had been achieved experimentally was in the complex mercury cuprate (138 K at ambient pressure \cite{PhysRevB.63.064511} and 166 K at high pressures \cite{0295-5075-72-3-458}).  
The new record of high T$_c$ was established for H$_3$S, a compound whose existence and superconductivity at 200 K were first predicted theoretically \cite{Duan2014} in 2014 using USPEX and then observed experimentally \cite{Drozdov2015} in 2015, and sparked a new wave of interest in hydrogen-rich superconductors.

In a previous theoretical study, 
Tse {\it et al.\/} reported a high-pressure metallic phase of SnH$_4$  with hexagonal P6/mmm symmetry group, which is a layered structure intercalated with H$_2$ units, and superconducts close to 80 K at 120 GPa \cite{Tse2007}.
Later, by using evolutionary algorithm, Gao {\it et al.\/} \cite{Gao2010} reported two novel low enthalpy metallic phases of SnH$_4$ with space groups P6$_3$/mmc and Ama2, which both have hexagonal layers of Sn atoms with semi-molecular H$_2$ units.  The reported stability ranges are 96-180 GPa for Ama2, and above 180 GPa for P6$_3$/mmc;
with T$_c$ values of 15-22 K at 120 GPa and 52-62 K at 200 GPa for Ama2 and  P6$_3$/mmc, respectively \cite{Gao2010}.

While SnH$_4$ was shown to be a relatively high-T$_c$ superconductor, other tin hydrides were not explored so far. At the same time, by now it is proven \cite{Zhang20122013} that totally unexpected compounds become stable under pressure, and that gives hope of finding even better superconductors. 
Hence, in this study, we systematically search for the stable compounds using the highly efficient method of variable-composition calculation (VCC). Apart from the previously reported phases of SnH$_4$, there is one metastable tetragonal phase of stannane with higher superconducting critical temperature. Other stable compounds e.g., SnH$_8$, SnH$_{12}$ and SnH$_{14}$ are found to form at high pressure with superconducting features.
 Moreover, we found a semi-molecular block of H$_3^{-}$ species in the I$\bar{4}$m2 structure of SnH$_8$. Novel H$_4^{-}$ is also observed in C2/m-SnH$_{12}$. We predict a high T$_c$ of 
81 K  at 220 GPa in the newly found compound SnH$_8$, 93 K for the phase SnH$_{12}$ at 250 GPa, 97 K for SnH$_{14}$ at 300 GPa and 91 K for the metastable phase of SnH$_4$ at 220 GPa.

\section*{Results}


Evolutionary variable-composition calculations up to 20 atoms in the unit cell were performed at 150, 200, 250 and 300 GPa. To further investigate the newly found compounds, fixed-composition structure predictions for the most promising compounds were performed, with one to three formula units per cell. Candidate low-enthalpy structures obtained by analysis of predicted structures are as follow: metastable I4/mmm-SnH$_4$, stable I$\bar{4}$m2-SnH$_8$, C2/m-SnH$_{12}$ and C2/m-SnH$_{14}$. 
In the I$\bar{4}$m2-SnH$_8$ structure predicted at 220 GPa, Sn atoms are packed between where H$_2$ and H$_3$ semi-molecules are located, in which the surrounding bent H$_3$ units are aligned perpendicular to one another with nearest distance of 1.347 \AA. In the C2/m-SnH$_{12}$, Sn atoms form a hexagonal packing intercalated with blocks of H$_2$ and H$_4$ semi-molecules.

\begin{figure*}[h]
\centerline{\includegraphics[width=1.0\textwidth]{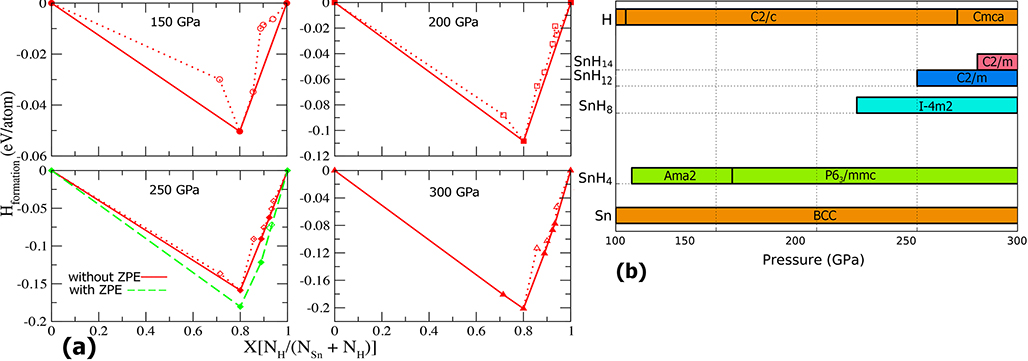}}
\caption{Thermodynamics of the Sn-H system. (a) Predicted formation enthalpy of Sn$_n$H$_m$ compounds with respect to decomposition into constituent elemental Sn and H$_2$ at different pressures. Red dashed lines connect data points, solid red lines denote the convex hull and green dashed line shows the effect of ZPE inclusion at 250 GPa. (b) Predicted pressure-composition phase diagram of Sn-H compounds.}\label{fig-1}
\end{figure*}

Figure \ref{fig-1}(a). shows the formation enthalpy ($\Delta H$) of Sn-H compounds at the chosen pressures. Significantly, in addition to reproducing various structures of solid SnH$_4$ \cite{Tse2007,Gao2010}, Sn \cite{Giefers2007} and H$_2$ \cite{Pickard2007}, novel compounds  SnH$_8$, SnH$_{12}$ and SnH$_{14}$ are found to be stable in a wide pressure range in our systematic evolutionary structure search.  
 It can be seen from Fig. \ref{fig-1}(a). that at around 200 GPa the tetragonal SnH$_8$ with the space group of I$\bar{4}$m2 lies above the tie-line, therefore, is metastable with respect to the decomposition to P6$_3$/mmc-SnH$_4$ and C2/c-H$_2$. Between 150 to 300 GPa, we predict stable phases of H$_2$, SnH$_4$, SnH$_8$, SnH$_{12}$, SnH$_{14}$ and Sn \cite{Giefers2007}. Some metastable forms of SnH$_6$, SnH$_9$ and SnH$_{16}$ are also shown in Fig. \ref{fig-1}(a). with open symbols.

\begin{figure*}[h]
\centerline{\includegraphics[width=\textwidth]{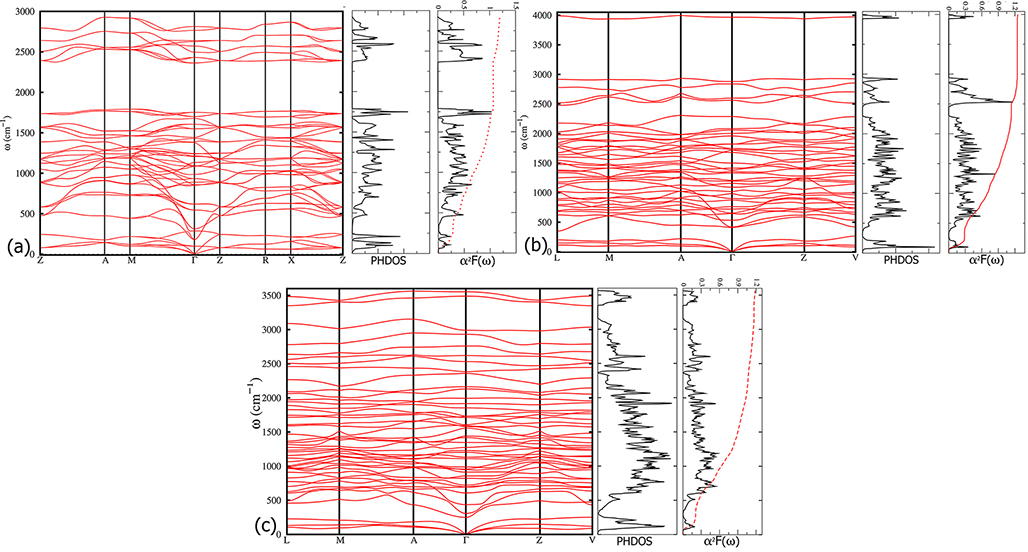}}
\caption{Phonon band structure, PHDOS, Eliashberg phonon spectral function $\alpha^2$F($\omega$) and electron-phonon integral $\lambda$($\omega$) of: (a) SnH$_8$ [I$\bar{4}$m2] at 220 GPa, (b) SnH$_{12}$ [C2/m] at 250 GPa and (c) SnH$_{14}$ [C2/m] at 300 GPa.}\label{lambda-phon-snh8}
\end{figure*}

SnH$_4$ is stable starting from 108 GPa as was predicted in previous report \cite{Gao2010}. It goes through a phase transition at 160 GPa.
 The higher pressure of 220 GPa favors  stabilization of SnH$_8$. SnH$_{12}$ and SnH$_{14}$ reach stability at the pressures of 250 GPa and 280 GPa, respectively, and remain stable at least up to 300 GPa. 
The structures of SnH$_n$ 
compounds are found to be dynamically stable within pressure ranges of their stability.
 Phonon band structures and phonon densities of states (PHDOS) are provided in Fig. \ref{lambda-phon-snh8}. In the I$\bar{4}$m2-SnH$_8$ structure, Sn atoms occupy the crystallographic 2a site and the H atoms are on the 4e, 8i and 4f sites (more structure information are provided in Table \ref{tab-1} ).



We checked the effects of zero-point energy using the quasi-harmonic approximation \cite{phonopy} at 250 GPa. The inclusion of zero-point noticeably lowered the formation enthalpy of SnH$_8$ with respect to SnH$_4$ and H$_2$ (Fig. \ref{fig-1}(a)), implying that this compound can be formed at lower pressure. Consequently, SnH$_{12}$ lies above the hull, suggesting more pressure is needed to stabilize C2/m-SnH$_{12}$. In accord with what we expect, zero-point energy does not change the topology of the phase diagram, but shifts transition pressures.

In I$\bar{4}$m2-SnH$_8$ structure, the ordered H atoms are split into two categories. One of H categories consists blocks of H$_3$ semi-molecules, which was previously observed in solid phases of BaH$_6$ \cite{Hooper2013}, in an unstable  structure of H$_5$Br ([H$_3$]Br[H$_2$])\cite{Duan}, and in an intriguing linear form of H$_3$ with the bond length of ~0.92 \AA\ and strong covalent bonds with ELF magnitude of ~0.8 in the H$_5$Te$_2$ \cite{Zhong}. 

 In contrast to H$_5$Br, which has approximately an equilateral triangle form of H$_3$, here we report the formation of H$_3$ in a bent molecule with the angle of 146.2$\degree$ and bond length of 0.867 \AA\ at 220 GPa in the I$\bar{4}$m2 structure. The other H category forms in semi-molecular H$_2$ units, which are aligned parallel to each other.

\begin{figure*}[h]
\centerline{\includegraphics[width=.6\textwidth]{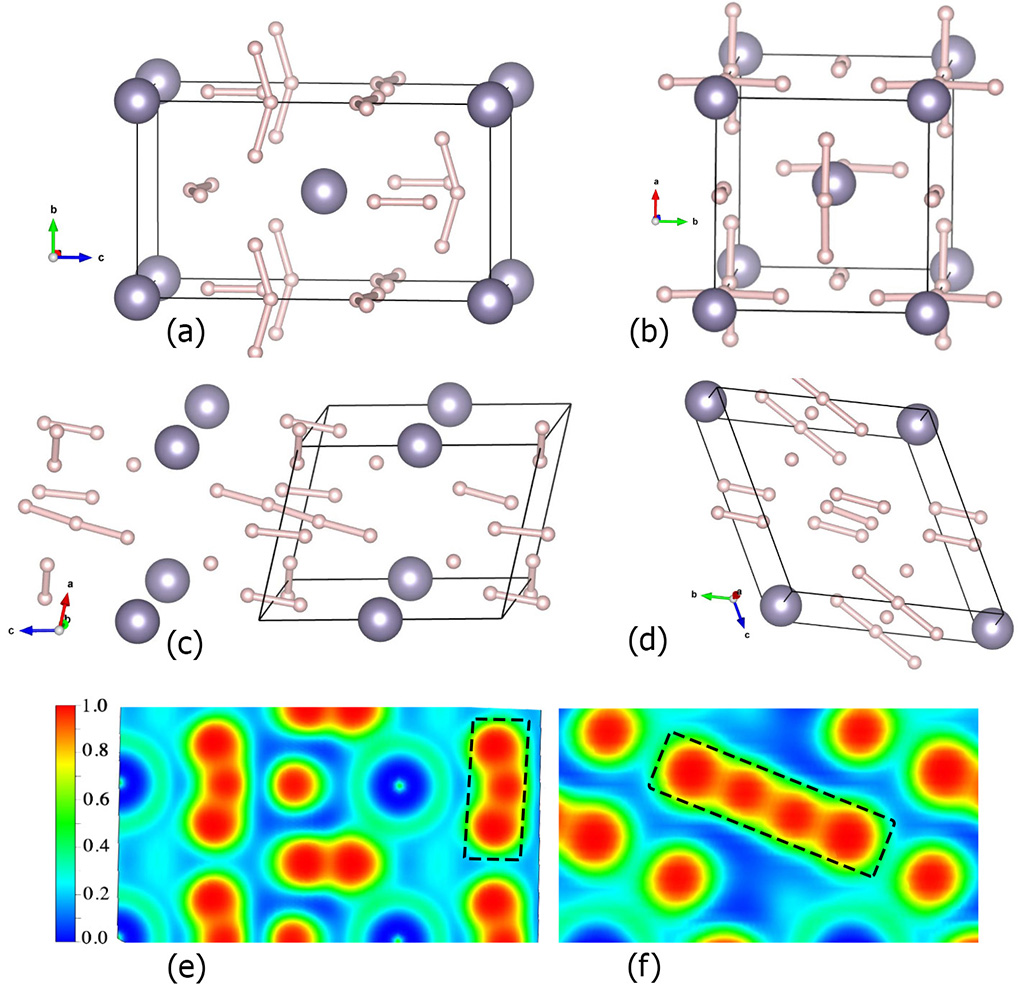}}
\caption{Predicted structures of (a),(b) SnH$_8$ [I$\bar{4}$m2], (c) SnH$_{12}$  [C2/m] and (d) SnH$_{14}$ [C2/m]. Large and small spheres represent Sn and H atoms, respectively. Electron localization functions (ELF) for (e) SnH$_8$ [I$\bar{4}$m2] at 220 GPa and (f) SnH$_{12}$ [C2/m] at 250 GPa.}\label{POS-final}
\end{figure*}

The presence of different  types of hydrogen can be explained based on charge transfer from Sn atom to each groups of hydrogen (H$_2$ or H$_3$).  I$\bar{4}$m2 structure consists of a [H$_2$][H$_3$]Sn[H$_3$] as shown in Fig. \ref{POS-final}(a),(b). The bond length in H$_3$ unit is 0.867 \AA, whereas H$_2$ has a longer bond length of 0.873 \AA. Contrary to isolated H$_2$ that already has a filled  $\sigma$ bond resulting in a strong covalent bond, in the H$_2$ and H$_3$ semi-molecules, anti-bonding electrons cause lengthening of the bond, subsequently resulting in a weaker covalent bond. 
The slightly longer H-H bond length compared to isolated H$_2$ molecule (0.74 \AA) is caused by charge transfer of 0.42 $\mathrm{e^-}$ and 0.48 $\mathrm{e^-}$ from Sn to each H$_2$ and H$_3$ units, respectively. Charge transfer is explained as an important factor in the formation of H$_2$ and H$_3$ blocks in the H$_4$Te, GeH$_4$, SnH$_4$, CaH$_6$, H$_5$Te$_2$, H$_5$Br, BaH$_6$ \cite{Hooper2013, Duan, Zhong,Wang2012, Gao2008}

Analysis of electron localization function (ELF) shows a high ELF value of 0.88 between H atoms within the unit, indicating strong covalent bonding features (Fig. \ref{POS-final}(e)). 
 Fig. \ref{POS-final}(e) indicated ELF magnitude of 0.37, pertaining to no covalent bond between Sn and hydrogen atoms. 

In C2/m-SnH$_{12}$, intriguing formation of novel H$_4$ semi-molecules is observed; at 250 GPa, the distance between blocks of H$_2$ is 0.9916 \AA. Higher pressure of 300 GPa decreases the distance to 0.8823 \AA, leading to a strong covalent bond in the H$_4^{-}$ units. Fig. \ref{POS-final}(f). demonstrates the strength of covalent bonds in the linear H$_4$ units with the ELF magnitude of 0.85.

\begin{figure*}[h]
\centerline{\includegraphics[width=.49\textwidth]{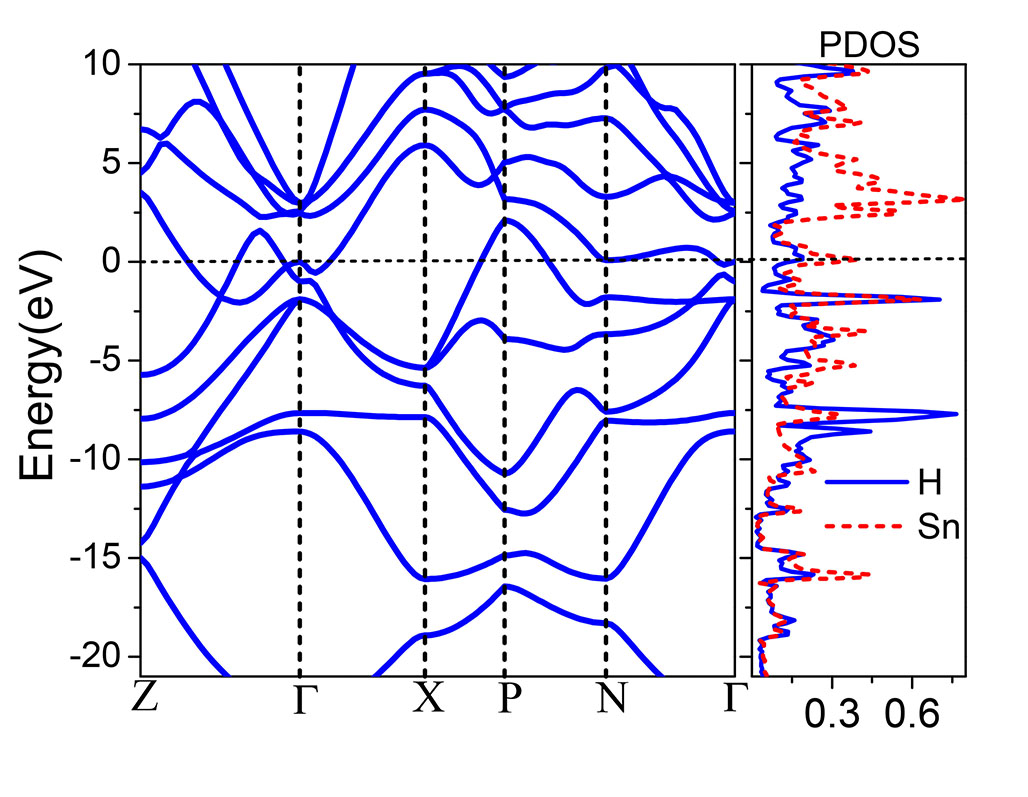}}
\caption{Electronic band structure and projected DOS on Sn and H atoms for SnH$_8$ [I$\bar{4}$m2] at 220 GPa.}\label{band-1-8}
\end{figure*}

Electronic band structure of I$\bar{4}$m2-SnH$_8$ is depicted in Fig. \ref{band-1-8}.  Occurrence of flat and steep bands near the Fermi level has been suggested as a condition for enhancing electron-phonon coupling and the formation of Cooper pairs. 

The calculated phonon dispersion curves and phonon density of states for I$\bar{4}$m2 structure of SnH$_8$ at 220 GPa is shown in Fig. \ref{lambda-phon-snh8}(a). Dynamical stability is clearly evidenced by the absence of any imaginary frequencies in the whole Brillouin zone. The low-frequency bands below 250 cm$^{-1}$ are mainly from the vibration of Sn atoms. 
 Modes between 300 and 1700 cm$^{-1}$ are mainly from the H-H wagging vibrations, and higher frequency vibrations above 2300 cm$^{-1}$ are due to H-H stretching vibrations in H$_2$ and H$_3$ units.

Low-frequency translational vibrations, mostly from Sn atom, contribute 23.7\% (9.2\%) to the total $\lambda$. Intermediate H-H wagging vibrations make a significant section of 65.7\% (67.9\%), and the rest is from stretching H vibrations, which contributes 10.6\% (22.9\%) for SnH$_8$ (SnH$_{12}$).
It is different from the superconductivity in the Cmcm-H$_2$Br\cite{Duan}, where Br translational vibrations contribute most to the total $\lambda$ and similar to the R$\bar{3}$m-H$_4$Te\cite{Zhong} and P4/mmm-BaH$_6$\cite{Hooper2013}, where medium-frequency H-wagging and bending modes contribute the most to the EPC.
In accord with our expectation, $\lambda$  increases almost linearly with the hydrogen content, 
 where 
 we found 60.2\%, 72.2\% and 77.1\%  contribution of H vibrations to the total $\lambda$ of SnH$_4$, SnH$_8$ and SnH$_{12}$, respectively. This denotes the dominant role of H in the superconductivity of H-rich compounds. 

The EPC calculations were performed to explore the superconductivity, in specific EPC constant $\lambda$, the Eliashberg phonon spectral function $\alpha^2F(\omega)$ and the logarithmic average of phonon frequencies $\omega_{log}$ (see Fig. \ref{lambda-phon-snh8}.). We can calculate T$_c$ based on the spectral function $\alpha^2$F($\omega$) and taking advantage of Allen-Dynes modified McMillan equation (Eq.~\ref{eq-allen}.) by using Coulomb pseudo-potential $\mu^*$ of 0.10 and 0.13 as widely accepted values (see Table \ref{tab-2}).


At 220 GPa, the predicted T$_c$ values for I$\bar{4}$m2-SnH$_8$ are 81 K and 72 K using $\mu^*$ values of 0.10 and 0.13, respectively.  The calculated T$_c$ slightly decreases with pressure (82 K at 200 GPa and 79 K at 300 GPa using $\mu^*$ = 0.10) with a pressure coefficient of -0.023 K/GPa ($\dfrac{dT_c}{dP}$). Reported $\lambda$ is comparable to the H$_3$Se ($\lambda$ = 1.09) at 200 GPa
\cite{Zhang}, but in I$\bar{4}$m2-SnH$_8$ structure, we have reduced $\langle\omega_{log}\rangle$ of 919 K (1477 K for H$_3$Se), resulting in a lower T$_c$ value.

In conclusion, we explored the energetically stable/metastable high-pressure phases of the Sn-H system in detail by means of {\it ab initio\/} evolutionary structure prediction. The results demonstrate that SnH$_8$, SnH$_{12}$ and SnH$_{14}$, reported for the first time in this work, are thermodynamically stable compounds that coexist stably with solid Sn, H$_2$ and SnH$_4$ in a wide pressure range starting from 220 to at least 300 GPa.

EPC calculations indicate that high-pressure SnH$_8$, SnH$_{12}$ and SnH$_{14}$ are phonon-mediated superconductors with T$_c$ values of 81.3, 93.2 and 97.2 K, respectively. T$_c$ decreases slightly with a rate of -0.023 K/GPa for I$\bar{4}$m2-SnH$_8$ structure. $\lambda$ is  high for SnH$_n$ compounds, comparable with H$_3$M-Im$\bar{3}$m, where M = S and Se\cite{Zhang}. 
Structures of SnH$_n$ compounds contain linear and non-linear H$_2^-$, H$_3^-$ and H$_4^-$ anions.    
Further experimental studies on the formation of SnH$_n$, n = 8, 12 and 14 at high pressure are needed, and these studies will serve as a guide for future experiments.

\section*{Methods}

To find stable and low-enthalpy metastable structures, we took advantage of evolutionary algorithm implemented in USPEX code \cite{Oganov2006,Glass2006, Oganov2011}, which has been extensively used to predict stable crystal structures with just a knowledge of the chemical composition and without any experimental information \cite{Martinez-Canales2009,Hu2013,Oganov2009}. 

In this method, the initial generation of structures and compositions is produced randomly with using random space group from the total list of 230 groups. 
 In order to find all stable stoichiometric compounds and the corresponding stable and metastable structures in the Sn-H binary system, we used VCC method implemented in USPEX \cite{Glass2006,Oganov2006}.

Structure relaxations were carried out using VASP package \cite{223}  in the framework of Density Functional Theory (DFT) adopting PBE-GGA (Perdew-Burke-Ernzerhof generalized gradient approximation) \cite{225}. The projector augmented-wave approach (PAW)  \cite{224} was used to describe the core electrons and their effects on valence orbitals.  The plane-wave kinetic energy cutoff was chosen as 1000 eV, and we used $\Gamma$-centered uniform k-points meshes to sample the Brillouin zone.

Phonons and thermodynamic properties of Sn-H compounds are calculated using the PHONOPY  package\cite{phonopy,phonopy1}. The supercell approach is used with simulation cell dimensions greater than 10 \AA\ (typically 3 $\times$ 3 $\times$ 3 representation of the primitive cell). 
We used valence electron configuration of 4d$^{10}$ 5s$^2$ 5p$^2$  and 1s$^1$ for tin and hydrogen, respectively. Phonon frequencies and electron-phonon coupling (EPC) coefficients are calculated using DFPT as implemented in the {\sc Quantum ESPRESSO} (QE) code \cite{227}.
In the QE calculations, we employed ultrasoft pseudopotentials along with PBE exchange and correlation functional\cite{225}. A plane-wave basis set with a cutoff of 70 Ry gave a convergence in energy with a precision of 1 meV/atom. 
The EPC parameter was calculated using 
4 $\times$ 4 $\times$ 3, 
5 $\times$ 5 $\times$ 4 and 
5 $\times$ 5 $\times$ 4 {\it q\/}-point meshes
 for I$\bar{4}$m2-SnH$_8$, C2/m-SnH$_{12}$ and C2/m-SnH$_{14}$, respectively.  
Denser {\it k\/}-point meshes, 
8 $\times$ 8 $\times$ 6, 
10 $\times$ 10 $\times$ 8 and 
10 $\times$ 10 $\times$ 8  
were used for convergence checks for the EPC parameter $\lambda$.
The superconducting T$_c$, was estimated using the Allen-Dynes modified McMillan equation \cite{PhysRevB.12.905}:
\begin{equation} 
T_c = \dfrac{\omega_{log}}{1.2} exp \left({\dfrac{-1.04 (1+ \lambda)}{\lambda - \mu^* ( 1 + 0.62 \lambda)}}\right)
\label{eq-allen} 
\end{equation} 

where $\omega_{log}$ is the logarithmic average frequency and $\mu^{*}$ is the Coulomb   pseudopotential, for which we used 0.10 and 0.13 values, which often give realistic results. The EPC constant and $\omega_{log}$ were calculated as: 
\begin{equation} 
\lambda = 2 \int_0^{\infty} \dfrac{\alpha^2 F (\omega)}{\omega} d\omega
\end{equation}
and
\begin{equation} 
\omega_{log} = exp \Big[ \dfrac{2}{\lambda} \int \dfrac{d\omega}{\omega} \alpha^2F(\omega) ln(\omega) \Big]
\end{equation}

\section*{Acknowledgements (not compulsory)}

We thank DARPA (Grants W31P4Q1210008 and W31P4Q1310005),
the Government of Russian Federation (14.A12.31.0003) and the
Foreign Talents Introduction and Academic Exchange Program
(B08040).
X.F.Z thanks the National Science Foundation of China (grant no. 11174152), the National 973 Program of China (grant no. 2012CB921900), and the Program for New Century Excellent Talents in University (grant no. NCET-12-0278). Calculations
were mainly performed on the cluster (QSH) in Oganov'	s lab at Stony Brook University.

\section*{Author contributions}

M.M.D.E. performed all the calculations presented in this article with help from Z.W., Q.Z. and H.D. Research was designed by A.R.O. S.W, M. R. and X-F. Z. analyzed data.  M.M.D.E., A.R.O. and Z.W. wrote the first draft of the paper and all authors contributed to revisions.

\section*{Additional information}

Competing financial interests: The authors declare no competing financial interests.

\newpage

\newpage

\begin{table}[ht]
\centering	
\caption{\bf Lattice parameters and atomic coordinates for SnH$_8$, SnH$_{12}$ and SnH$_{14}$ at 220, 250 and 300 GPa, respectively.}
\begin{tabular*}{10cm}{@{\extracolsep{\fill}}cccccc}
 \hline\hline
 Space group & Lattice & Atom & x & y & z \\ [0.5ex] 
             &parameters &   &   &   \\ [0.5ex] 
 \hline
 I$\bar{4}$m2 & a=3.076 \AA\ & Sn(2a)    & 0.0000    & 0.0000 & 0.0000 \\ 
   SnH$_8$    & c=5.523 \AA\ & H$_1$(8i) & 0.2729    & 0.0000 & 0.3331 \\ 
              & at 220 GPa   & H$_2$(4e) & 0.0000    & 0.0000 & 0.6208 \\
              &              & H$_3$(4f) & 0.0000    & 0.5000 & 0.1701 \\
              
 \hline
  C2/m        & a=5.191 \AA\      & Sn(2d)     & 0.0000 & 0.5000 & 0.5000 \\ 
  SnH$_{12}$  & b=3.065 \AA\      & H$_1$(4i)  & 0.0495 & 0.0000 & 0.6553 \\
              & c=7.388  \AA\     & H$_2$(4i)  & 0.4564 & 0.0000 & 0.7226 \\
  & $\beta$=148.95$^{\circ}$      & H$_3$(4i)  & 0.3428 & 0.0000 & 0.8832 \\
              &    at 250 GPa     & H$_4$(8i)  & 0.3810 & 0.2399 & 0.1123 \\
              &                   & H$_5$(4g)  & 0.0000 & 0.1233 & 0.0000 \\ 
 \hline
 C2/m         & a=7.129   \AA\         & Sn(2b)     & 0.0000 & 0.5000 & 0.0000 \\ 
 SnH$_{14}$     & b=2.730 \AA\         & H$_1$(4i)  & 0.3651 & 0.0000 & 0.7031 \\
              & c=3.673    \AA\        & H$_2$(4i)  & 0.1857 & 0.0000 & 0.9852 \\
              & $\beta$=60.71$^{\circ}$& H$_3$(4i)  & 0.0732 & 0.0000 & 0.6252 \\
              &                        & H$_4$(4i)  & 0.8063 & 0.0000 & 0.8090 \\
              &    at 300 GPa          & H$_5$(8i)  & 0.2365 & 0.2808 & 0.4035 \\
              &                        & H$_6$(2d)  & 0.0000 & 0.5000 & 0.5000 \\ 
              &                        & H$_7$(2c)  & 0.0000 & 0.0000 & 0.5000 \\     
 \hline
\end{tabular*}
 \label{tab-1}
\end{table}

\newpage

\begin{table}[ht]
\centering
\caption{\bf The calculated EPC parameter ($\lambda$), logarithmic average phonon frequency ($\omega_{log}$) and critical temperature (T$_c$) (with $\mu^*$ = 0.10 and 0.13) for metastable SnH$_4$, stable SnH$_8$, SnH$_{12}$ and SnH$_{14}$ at 220, 220, 250 and 300 GPa, respectively.}
 \begin{tabular*}{10cm}{@{\extracolsep{\fill}}ccccc}
 \hline\hline
 Structure & Pressure (GPa) & $\lambda$ & $\omega_{log}$ (K) & T$_c$ (K) \\ [0.5ex] 
 \hline
 I4/mmm      & 220 & 1.180    & 1025  & 91 ($\mu^*$=0.10)  \\
      SnH$_4$              &     &         &        & 80  ($\mu^*$=0.13) \\

 \hline
 I$\bar{4}$m2-SnH$_8$ & 220 & 1.188   & 919  & 81 ($\mu^*$=0.10)   \\ 
                      &   &   &   &  72 ($\mu^*$=0.13) \\

 \hline
  C2/m-SnH$_{12}$ & 250 & 1.250    & 991 & 93 ($\mu^*$=0.10) \\
                  &     &         &        & 83 ($\mu^*$=0.13) \\
 \hline
 C2/m-SnH$_{14}$  & 300 & 1.187    &  1099 & 97 ($\mu^*$=0.10)\\ 
                  &     &       &    & 86 ($\mu^*$=0.13) \\
 \hline
 \label{tab-2}
\end{tabular*}
\end{table}


\begin{thebibliography}{10}
\expandafter\ifx\csname url\endcsname\relax
  \def\url#1{\texttt{#1}}\fi
\expandafter\ifx\csname urlprefix\endcsname\relax\def\urlprefix{URL }\fi
\providecommand{\bibinfo}[2]{#2}
\providecommand{\eprint}[2][]{\url{#2}}

\bibitem{Nellis2006}
\bibinfo{author}{Nellis, W.~J.}
\newblock \bibinfo{title}{Dynamic compression of materials: metallization of
  fluid hydrogen at high pressures}.
\newblock \emph{\bibinfo{journal}{Rep. Prog. Phys.}}
  \textbf{\bibinfo{volume}{69}}, \bibinfo{pages}{1479} (\bibinfo{year}{2006}).

\bibitem{Loubeyre2002}
\bibinfo{author}{Loubeyre, P.}, \bibinfo{author}{Occelli, F.} \&
  \bibinfo{author}{LeToullec, R.}
\newblock \bibinfo{title}{{Optical studies of solid hydrogen to 320 GPa and
  evidence for black hydrogen.}}
\newblock \emph{\bibinfo{journal}{Nature}} \textbf{\bibinfo{volume}{416}},
  \bibinfo{pages}{613--617} (\bibinfo{year}{2002}).

\bibitem{Zha2013}
\bibinfo{author}{Zha, C.-S.}, \bibinfo{author}{Liu, Z.},
  \bibinfo{author}{Ahart, M.}, \bibinfo{author}{Boehler, R.} \&
  \bibinfo{author}{Hemley, R.~J.}
\newblock \bibinfo{title}{{High-Pressure Measurements of Hydrogen Phase IV
  Using Synchrotron Infrared Spectroscopy}}.
\newblock \emph{\bibinfo{journal}{Phys. Rev. Lett.}}
  \textbf{\bibinfo{volume}{110}}, \bibinfo{pages}{217402}
  (\bibinfo{year}{2013}).

\bibitem{Howie2012}
\bibinfo{author}{Howie, R.~T.}, \bibinfo{author}{Scheler, T.},
  \bibinfo{author}{Guillaume, C.~L.} \& \bibinfo{author}{Gregoryanz, E.}
\newblock \bibinfo{title}{{Proton tunneling in phase IV of hydrogen and
  deuterium}}.
\newblock \emph{\bibinfo{journal}{Phys. Rev. B}} \textbf{\bibinfo{volume}{86}},
  \bibinfo{pages}{214104} (\bibinfo{year}{2012}).

\bibitem{Review2014}
\bibinfo{author}{Zha, C.-S.}, \bibinfo{author}{Cohen, R.~E.},
  \bibinfo{author}{Mao, H.-k.} \& \bibinfo{author}{Hemley, R.~J.}
\newblock \bibinfo{title}{{Raman measurements of phase transitions in dense
  solid hydrogen and deuterium to 325 GPa}}.
\newblock \emph{\bibinfo{journal}{Proc. Natl. Acad. Sci.}}
  \textbf{\bibinfo{volume}{111}}, \bibinfo{pages}{4792--4797}
  (\bibinfo{year}{2014}).

\bibitem{Ashcroft1968}
\bibinfo{author}{Ashcroft, N.~W.}
\newblock \bibinfo{title}{Metallic hydrogen: A high-temperature
  superconductor?}
\newblock \emph{\bibinfo{journal}{Phys. Rev. Lett.}}
  \textbf{\bibinfo{volume}{21}}, \bibinfo{pages}{1748--1749}
  (\bibinfo{year}{1968}).

\bibitem{Cudazzo2010}
\bibinfo{author}{Cudazzo, P.} \emph{et~al.}
\newblock \bibinfo{title}{{Electron-phonon interaction and superconductivity in
  metallic molecular hydrogen. II. Superconductivity under pressure}}.
\newblock \emph{\bibinfo{journal}{Phys. Rev. B}} \textbf{\bibinfo{volume}{81}},
  \bibinfo{pages}{134506} (\bibinfo{year}{2010}).

\bibitem{Cudazzo2010a}
\bibinfo{author}{Cudazzo, P.} \emph{et~al.}
\newblock \bibinfo{title}{{Electron-phonon interaction and superconductivity in
  metallic molecular hydrogen. I. Electronic and dynamical properties under
  pressure}}.
\newblock \emph{\bibinfo{journal}{Phys. Rev. B}} \textbf{\bibinfo{volume}{81}},
  \bibinfo{pages}{134505} (\bibinfo{year}{2010}).

\bibitem{Ashcroft2004}
\bibinfo{author}{Ashcroft, N.~W.}
\newblock \bibinfo{title}{Hydrogen dominant metallic alloys: High temperature
  superconductors?}
\newblock \emph{\bibinfo{journal}{Phys. Rev. Lett.}}
  \textbf{\bibinfo{volume}{92}}, \bibinfo{pages}{187002}
  (\bibinfo{year}{2004}).

\bibitem{Eremets2008}
\bibinfo{author}{Eremets, M.~I.}, \bibinfo{author}{Trojan, I.~A.},
  \bibinfo{author}{Medvedev, S.~A.}, \bibinfo{author}{Tse, J.~S.} \&
  \bibinfo{author}{Yao, Y.}
\newblock \bibinfo{title}{Superconductivity in hydrogen dominant materials:
  Silane}.
\newblock \emph{\bibinfo{journal}{Science}} \textbf{\bibinfo{volume}{319}},
  \bibinfo{pages}{1506--1509} (\bibinfo{year}{2008}).

\bibitem{Zhang}
\bibinfo{author}{{Zhang}, S.} \emph{et~al.}
\newblock \bibinfo{title}{{Phase Diagram and High-Temperature Superconductivity
  of Compressed Selenium Hydrides}}.
\newblock \emph{\bibinfo{journal}{ArXiv e-prints}}  (\bibinfo{year}{2015}).
\newblock \eprint{1502.02607}.

\bibitem{Wang2012}
\bibinfo{author}{Wang, H.}, \bibinfo{author}{Tse, J.~S.},
  \bibinfo{author}{Tanaka, K.}, \bibinfo{author}{Iitaka, T.} \&
  \bibinfo{author}{Ma, Y.}
\newblock \bibinfo{title}{{Superconductive sodalite-like clathrate calcium
  hydride at high pressures}}.
\newblock \emph{\bibinfo{journal}{Proc. Natl. Acad. Sci.}}
  \textbf{\bibinfo{volume}{109}}, \bibinfo{pages}{6463--6466}
  (\bibinfo{year}{2012}).

\bibitem{Gao2010}
\bibinfo{author}{Gao, G.} \emph{et~al.}
\newblock \bibinfo{title}{{High-pressure crystal structures and
  superconductivity of Stannane (SnH4)}}.
\newblock \emph{\bibinfo{journal}{Proc. Natl. Acad. Sci.}}
  \textbf{\bibinfo{volume}{107}}, \bibinfo{pages}{1317--1320}
  (\bibinfo{year}{2010}).

\bibitem{PhysRevB.84.054543}
\bibinfo{author}{Zhou, X.-F.} \emph{et~al.}
\newblock \bibinfo{title}{Superconducting high-pressure phase of platinum
  hydride from first principles}.
\newblock \emph{\bibinfo{journal}{Phys. Rev. B}} \textbf{\bibinfo{volume}{84}},
  \bibinfo{pages}{054543} (\bibinfo{year}{2011}).

\bibitem{Hu2013}
\bibinfo{author}{Hu, C.-H.} \emph{et~al.}
\newblock \bibinfo{title}{{Pressure-Induced Stabilization and
  Insulator-Superconductor Transition of BH}}.
\newblock \emph{\bibinfo{journal}{Phys. Rev. Lett.}}
  \textbf{\bibinfo{volume}{110}}, \bibinfo{pages}{165504}
  (\bibinfo{year}{2013}).

\bibitem{Duan2014}
\bibinfo{author}{Duan, D.} \emph{et~al.}
\newblock \bibinfo{title}{{Pressure-induced metallization of dense (H2S)2H2
  with high-Tc superconductivity.}}
\newblock \emph{\bibinfo{journal}{Sci. Rep.}} \textbf{\bibinfo{volume}{4}},
  \bibinfo{pages}{6968} (\bibinfo{year}{2014}).

\bibitem{Gao2008}
\bibinfo{author}{Gao, G.} \emph{et~al.}
\newblock \bibinfo{title}{Superconducting high pressure phase of germane}.
\newblock \emph{\bibinfo{journal}{Phys. Rev. Lett.}}
  \textbf{\bibinfo{volume}{101}}, \bibinfo{pages}{107002}
  (\bibinfo{year}{2008}).

\bibitem{PhysRevB.63.064511}
\bibinfo{author}{Lokshin, K.~A.} \emph{et~al.}
\newblock \bibinfo{title}{{Enhancement of ${T}_{c}$ in
  ${\mathrm{HgBa}}_{2}{\mathrm{Ca}}_{2}{\mathrm{Cu}}_{3}{\mathrm{O}}_{8+\ensuremath{\delta}}$
  by fluorination}}.
\newblock \emph{\bibinfo{journal}{Phys. Rev. B}} \textbf{\bibinfo{volume}{63}},
  \bibinfo{pages}{064511} (\bibinfo{year}{2001}).

\bibitem{0295-5075-72-3-458}
\bibinfo{author}{Monteverde, M.} \emph{et~al.}
\newblock \bibinfo{title}{{High-pressure effects in fluorinated
  ${\mathrm{HgBa}}_{2}{\mathrm{Ca}}_{2}{\mathrm{Cu}}_{3}{\mathrm
  {O}}_{8+\ensuremath{\delta}}$}}.
\newblock \emph{\bibinfo{journal}{Europhys. Lett.}}
  \textbf{\bibinfo{volume}{72}}, \bibinfo{pages}{458} (\bibinfo{year}{2005}).

\bibitem{Drozdov2015}
\bibinfo{author}{Drozdov, A.~P.}, \bibinfo{author}{Eremets, M.~I.},
  \bibinfo{author}{Troyan, I.~A.}, \bibinfo{author}{Ksenofontov, V.} \&
  \bibinfo{author}{Shylin, S.~I.}
\newblock \bibinfo{title}{{Conventional superconductivity at 203 kelvin at high
  pressures in the sulfur hydride system}}.
\newblock \emph{\bibinfo{journal}{Nature}} \textbf{\bibinfo{volume}{525}},
  \bibinfo{pages}{73--76} (\bibinfo{year}{2015}).

\bibitem{Tse2007}
\bibinfo{author}{Tse, J.~S.}, \bibinfo{author}{Yao, Y.} \&
  \bibinfo{author}{Tanaka, K.}
\newblock \bibinfo{title}{{Novel Superconductivity in Metallic SnH4 under High
  Pressure}}.
\newblock \emph{\bibinfo{journal}{Phys. Rev. Lett.}}
  \textbf{\bibinfo{volume}{98}}, \bibinfo{pages}{117004}
  (\bibinfo{year}{2007}).

\bibitem{Zhang20122013}
\bibinfo{author}{Zhang, W.} \emph{et~al.}
\newblock \bibinfo{title}{Unexpected stable stoichiometries of sodium
  chlorides}.
\newblock \emph{\bibinfo{journal}{Science}} \textbf{\bibinfo{volume}{342}},
  \bibinfo{pages}{1502--1505} (\bibinfo{year}{2013}).

\bibitem{Giefers2007}
\bibinfo{author}{Giefers, H.} \emph{et~al.}
\newblock \bibinfo{title}{{Phonon Density of States of Metallic Sn at High
  Pressure}}.
\newblock \emph{\bibinfo{journal}{Phys. Rev. Lett.}}
  \textbf{\bibinfo{volume}{98}}, \bibinfo{pages}{245502}
  (\bibinfo{year}{2007}).

\bibitem{Pickard2007}
\bibinfo{author}{Pickard, C.~J.} \& \bibinfo{author}{Needs, R.~J.}
\newblock \bibinfo{title}{{Structure of phase III of solid hydrogen}}.
\newblock \emph{\bibinfo{journal}{Nature Phys.}} \textbf{\bibinfo{volume}{3}},
  \bibinfo{pages}{473--476} (\bibinfo{year}{2007}).

\bibitem{phonopy}
\bibinfo{author}{Togo, A.}, \bibinfo{author}{Oba, F.} \&
  \bibinfo{author}{Tanaka, I.}
\newblock \bibinfo{title}{{First-principles calculations of the ferroelastic
  transition between rutile-type and ${\text{CaCl}}_{2}$-type
  ${\text{SiO}}_{2}$ at high pressures}}.
\newblock \emph{\bibinfo{journal}{Phys. Rev. B}} \textbf{\bibinfo{volume}{78}},
  \bibinfo{pages}{134106} (\bibinfo{year}{2008}).

\bibitem{Hooper2013}
\bibinfo{author}{Hooper, J.}, \bibinfo{author}{Altintas, B.},
  \bibinfo{author}{Shamp, A.} \& \bibinfo{author}{Zurek, E.}
\newblock \bibinfo{title}{{Polyhydrides of the alkaline earth metals: A look at
  the extremes under pressure}}.
\newblock \emph{\bibinfo{journal}{J. Phys. Chem. C}}
  \textbf{\bibinfo{volume}{117}}, \bibinfo{pages}{2982--2992}
  (\bibinfo{year}{2013}).

\bibitem{Duan}
\bibinfo{author}{{Duan}, D.} \emph{et~al.}
\newblock \bibinfo{title}{{Decomposition of solid hydrogen bromide at high
  pressure}}.
\newblock \emph{\bibinfo{journal}{ArXiv e-prints}}  (\bibinfo{year}{2015}).
\newblock \eprint{1504.01196}.

\bibitem{Zhong}
\bibinfo{author}{{Zhong}, X.} \emph{et~al.}
\newblock \bibinfo{title}{{Tellurium Hydrides at High Pressures:
  High-temperature Superconductors}}.
\newblock \emph{\bibinfo{journal}{ArXiv e-prints}}  (\bibinfo{year}{2015}).
\newblock \eprint{1503.00396}.

\bibitem{Oganov2006}
\bibinfo{author}{Oganov, A.~R.} \& \bibinfo{author}{Glass, C.~W.}
\newblock \bibinfo{title}{{Crystal structure prediction using ab initio
  evolutionary techniques: principles and applications.}}
\newblock \emph{\bibinfo{journal}{J. Chem. Phys}}
  \textbf{\bibinfo{volume}{124}}, \bibinfo{pages}{244704}
  (\bibinfo{year}{2006}).

\bibitem{Glass2006}
\bibinfo{author}{Glass, C.~W.}, \bibinfo{author}{Oganov, A.~R.} \&
  \bibinfo{author}{Hansen, N.}
\newblock \bibinfo{title}{{USPEX - Evolutionary crystal structure prediction}}.
\newblock \emph{\bibinfo{journal}{Comp. Phys. Comm.}}
  \textbf{\bibinfo{volume}{175}}, \bibinfo{pages}{713--720}
  (\bibinfo{year}{2006}).

\bibitem{Oganov2011}
\bibinfo{author}{Oganov, A.~R.}, \bibinfo{author}{Lyakhov, A.~O.} \&
  \bibinfo{author}{Valle, M.}
\newblock \bibinfo{title}{{How Evolutionary Crystal Structure Prediction
  Works-and Why}}.
\newblock \emph{\bibinfo{journal}{Acc. Chem. Res.}}
  \textbf{\bibinfo{volume}{44}}, \bibinfo{pages}{227--237}
  (\bibinfo{year}{2011}).

\bibitem{Martinez-Canales2009}
\bibinfo{author}{Martinez-Canales, M.} \emph{et~al.}
\newblock \bibinfo{title}{{Novel Structures and Superconductivity of Silane
  under Pressure}}.
\newblock \emph{\bibinfo{journal}{Phys. Rev. Lett.}}
  \textbf{\bibinfo{volume}{102}} (\bibinfo{year}{2009}).

\bibitem{Oganov2009}
\bibinfo{author}{Oganov, A.~R.} \emph{et~al.}
\newblock \bibinfo{title}{{Ionic high-pressure form of elemental boron}}.
\newblock \emph{\bibinfo{journal}{Nature}} \textbf{\bibinfo{volume}{457}},
  \bibinfo{pages}{863--867} (\bibinfo{year}{2009}).

\bibitem{223}
\bibinfo{author}{Kresse, G.} \& \bibinfo{author}{Furthm\"uller, J.}
\newblock \bibinfo{title}{Efficient iterative schemes for \textit{ab initio}
  total-energy calculations using a plane-wave basis set}.
\newblock \emph{\bibinfo{journal}{Phys. Rev. B}} \textbf{\bibinfo{volume}{54}},
  \bibinfo{pages}{11169--11186} (\bibinfo{year}{1996}).

\bibitem{225}
\bibinfo{author}{Perdew, J.~P.}, \bibinfo{author}{Burke, K.} \&
  \bibinfo{author}{Ernzerhof, M.}
\newblock \bibinfo{title}{Generalized gradient approximation made simple}.
\newblock \emph{\bibinfo{journal}{Phys. Rev. Lett.}}
  \textbf{\bibinfo{volume}{77}}, \bibinfo{pages}{3865--3868}
  (\bibinfo{year}{1996}).

\bibitem{224}
\bibinfo{author}{Bl\"ochl, P.~E.}
\newblock \bibinfo{title}{Projector augmented-wave method}.
\newblock \emph{\bibinfo{journal}{Phys. Rev. B}} \textbf{\bibinfo{volume}{50}},
  \bibinfo{pages}{17953--17979} (\bibinfo{year}{1994}).

\bibitem{phonopy1}
\bibinfo{author}{Togo, A.} \& \bibinfo{author}{Tanaka, I.}
\newblock \bibinfo{title}{First principles phonon calculations in materials
  science}.
\newblock \emph{\bibinfo{journal}{Scr. Mater.}} \textbf{\bibinfo{volume}{108}},
  \bibinfo{pages}{1--5} (\bibinfo{year}{2015}).

\bibitem{227}
\bibinfo{author}{Giannozzi, P.} \emph{et~al.}
\newblock \bibinfo{title}{Quantum espresso: a modular and open-source software
  project for quantum simulations of materials}.
\newblock \emph{\bibinfo{journal}{J. Phys. Condens. Matter}}
  \textbf{\bibinfo{volume}{21}}, \bibinfo{pages}{395502}
  (\bibinfo{year}{2009}).

\bibitem{PhysRevB.12.905}
\bibinfo{author}{Allen, P.~B.} \& \bibinfo{author}{Dynes, R.~C.}
\newblock \bibinfo{title}{Transition temperature of strong-coupled
  superconductors reanalyzed}.
\newblock \emph{\bibinfo{journal}{Phys. Rev. B}} \textbf{\bibinfo{volume}{12}},
  \bibinfo{pages}{905--922} (\bibinfo{year}{1975}).

\end{thebibliography}
\end{document}